\documentstyle[prd,aps,floats,epsfig]{revtex}  
\input epsf.tex  
\def\DESepsf(#1 width #2){\epsfxsize=#2 \epsfbox{#1}}  
\def\xgq{x_{\tilde g \tilde q}}
\def\msq{\widetilde m}
\def\mgl{m_{\tilde g}}
\def\pbslash{\not\!{p_b}}
\def\pdslash{\not\!{p_d}}
\def\qslash{\not\!{q}}
\begin{document}  
%  
%\preprint{\vbox{\hbox{}}}  
\draft  
\twocolumn[\hsize\textwidth\columnwidth\hsize\csname
@twocolumnfalse\endcsname
\title{Supersymmetric Model Contributions to $B^0_d$--$\bar B^0_d$ Mixing and
$B\to\pi\pi,\,\rho\gamma$ Decays}  
%
%\thanks{Email address: ckchua@phys.ntu.edu.tw}   
%
\author{Abdesslam Arhrib, Chun-Khiang Chua and Wei-Shu Hou\\}  
\address{  
Department of Physics, National Taiwan University,   
Taipei, Taiwan 10764, R.O.C.  
}  
\date{July 7, 2001}  
\maketitle  
\begin{abstract}  
Recent results from Belle and BaBar Collaborations hint at
a small $\sin2\phi_1$, 
while the measured $B\to\pi\pi$ rate also
seems to be on the low side. 
Supersymmetric (SUSY) models with down squark
mixings can account for the deficits in both cases.
By studying the origin of SUSY contributions that could impact
on $B^0_d$--$\bar B^0_d$ mixing and $B\to\pi\pi$ decay,
we find that the former would most likely arise from left-left or 
right-right squark mixings, 
while the latter would come from left-right squark mixings.
These two processes in general are not much correlated in the
Minimum Supersymmetric Standard Model. 
If the smallness of $B\to\pi\pi$ is due to SUSY models,
one would likely have large $B\to\rho\gamma$ from chiral enhancement,
and the rate could be within present experimental reach.
Even if $B\to\rho\gamma$ is not greatly enhanced,
it could have large mixing dependent CP violation. 
\end{abstract}
]  
%  
%\preprint{\vbox{\hbox{}}}  
%  
%\preprint{\vbox{\hbox{}}}  
%  
%\preprint{\vbox{\hbox{}}}  
%  
%\maketitle  
%  
\section{Introduction}
The $CP$ asymmetries in $B^0\rightarrow J/\psi K_S$, $J/\psi K_L$
decays have been studied by several experimental groups 
\cite{CDF,OPAL,ALEPH,Belle,BaBar}.  
It is well known that one of the phase angles
of the Standard Model (SM) unitarity triangle, $\sin2\phi_1$,
can be measured via the asymmetry,  
\begin{eqnarray}  
a_{J/\psi K_S}&=&{\Gamma (B^0(t)\rightarrow J/\psi K^0_S)-  
                  \Gamma (\bar B^0(t)\rightarrow J/\psi K^0_S) \over  
                        \Gamma (B^0(t)\rightarrow J/\psi K^0_S)+  
                        \Gamma (\bar B^0(t)\rightarrow J/\psi K^0_S)}
\nonumber \\  
               &=&-\sin 2\phi_1%\,
                  \sin \Delta m_{B_d} t.  
\end{eqnarray}  
The CDF Collaboration finds  
$\sin 2 \phi_1=0.79{+0.41\atop-0.44}$ \cite{CDF}  
with Tevatron Run-I data,  
while the OPAL and ALEPH Collaborations give  
$\sin 2 \phi_1=3.2{+1.8\atop -2.0}\pm 0.5$\cite{OPAL},
$0.84{+0.82\atop -1.04}\pm 0.16$\cite{ALEPH}, respectively.  
Recently, however, the BaBar and Belle Collaborations announced their 
results on the measurement of this asymmetry.
The Belle Collaboration reports
$\sin 2 \phi_1=0.58{+0.32+0.09\atop-0.34-0.10}$\cite{Belle},
%where the central value is on the small side,
while the BaBar Collaboration gives the even smaller
$\sin 2 \phi_1=0.34\pm0.20\pm0.05$ \cite{BaBar}.
%which is even smaller. 
When combined with previous CDF and LEP results, 
the average value is
$\sin 2 \phi_1=0.48{\pm 0.16}$.
While this is consistent with 
the Cabibbo-Kobayashi-Maskawa (CKM) fit value of 
$\sin 2\phi_1=0.698\pm0.066$\cite{Ciuchini} or 
$0.47\leq\sin 2 \phi_1\leq0.93$ (95\%C.L.) \cite{Ali2000}, 
the central value is rather small.
It could be hinting at the presence of new physics effects,
especially if the value persists. 
In this case, we may need a large new physics contribution \cite{ChuaHou}
in $B^0$--$\bar B^0$ mixing, comparable to the SM amplitude, 
to account for the smallness of $a_{J/\psi K_S}$.
This is because it is very hard for new physics to affect 
the Cabibbo favored $b\to c\bar c s$ decay amplitude.

The first result on the charmless decay mode $B^0\to\pi^+\pi^-$
was given by the CLEO Collaboration, giving 
Br$(B^0\to\pi^+\pi^-)=(4.3{+1.6\atop-1.4}\pm0.5)\times10^{-6}$\cite{CLEOpipi}.
BaBar and Belle Collaborations also reported recently their results,
Br$(B^0\to\pi^+\pi^-)=(4.1{\pm1.0\pm0.7})\times10^{-6}$\cite{BaBarpipi},
$(5.9{+2.4\atop-2.1}\pm0.5)\times10^{-6}$\cite{Bellepipi}, respectively.
Note that the BaBar and Belle measurements are all lower than
their reported results at summer 2000 conferences \cite{BaBarpipiichep,Bellepipiichep}.  
The combined result with averaged Br$(B^0\to\pi^+\pi^-)=4.4\pm0.9$
seem to be on the low side when compared to the SM prediction
using factorization approach, 
Br$(B^0\to\pi^+\pi^-)\sim 10\times10^{-6}$\cite{Ali},
for $\phi_3\sim 60^\circ$,
and remains true when compared to the QCD factorization
%Beneke-Buchalla-Neubert-Sachrajda (BBNS)
result \cite{BBNS} of Br$(B^0\to\pi^+\pi^-)\sim 8\times10^{-6}$.
In a recent work on QCD factorization, Br$(\pi^+\pi^-)$ can be 
close to the experimental value,
however, the lowness of the rate
would subject the SM to considerable stress 
\cite{Beneke:2001ev}. 
A 20\%--40\% or more reduction in the branching ratio is welcome. 
In SM, the tree amplitude dominates over the penguin amplitude, 
which is about 30\% of the former.
Thus we may need large contribution from New Physics 
if it is responsible for the smallness of the rate.

We have two cases where we are in a situation that New Physics
contributions should be large,
if it is responsible for the smallness of measurements.
As one of the leading candidates for new physics, supersymmetry (SUSY)  
helps resolve many of the potential problems that emerge when one   
extends beyond the SM, for example the gauge hierarchy problem,   
unification of SU(3)$\times $SU(2)$\times $U(1) gauge couplings,   
and so on \cite{SUSY}.
In the context of SUSY, we then ask the following questions~:
Is it possible for SUSY models to account for the smallness in both processes?
If so, are they correlated, since both of them are $b\to d$ flavor changing processes?
Since New Physics contributions would be large, can we find other 
related effects? 

To analyse SUSY contributions,
we follow the approach of Ref. \cite{Gabbiani}.
As will be discussed later, gluino exchange diagrams
induced by $\tilde d$--$\tilde b$ mixings give 
dominant contributions in both of the above mention processes.
We do not aim at constructing any explicit models,
hence we have not considered other flavor changing processes 
such as $K^0$-$\bar K^0$ mixing, $D^0$-$\bar D^0$ mixing, 
Br$(B\to X_s\gamma)$ and the neutron electric dipole moment, etc., 
since these are controlled by other parameters.
Our strategy has been simply to study the implications on
$\tilde d$-$\tilde b$ squark mixings from new data on
$B_d$ mixing and $B\to \pi\pi$ decay,
and make inference on other modes, 
such as $B\to \rho\gamma$ which is quite correlated
with effects in $B\to \pi\pi$.
We have assumed that models can be constructed such that
SUSY can impact on the modes considered here,
but do not run into trouble with 
other stringent low energy constraints 
(see e.g. Ref. \cite{ChuaHou} for $B_d$ mixing case).

We organize this paper in the following way:
We discuss SUSY contribution to $B^0$--$\bar B^0$ mixing, 
$B\to\pi\pi$ and radiative B decays in the next two sections.
We then give some discussion, followed by conclusion in the last section.

\section{$B^0$--$\bar B^0$ mixing in SUSY models}

The effective Hamiltonian for $B^0_d$--$\bar B^0_d$ mixing  
from SUSY contributions is given by   
\begin{equation}  
\label{Heff}  
H_{\rm eff}=-{\sum_i C_i {\cal O}_i},   
\end{equation}  
where,  
\begin{eqnarray}  
&{\cal O}_1=\bar d^\alpha_{L}\gamma_\mu b^\alpha_L\,  
           \bar d^\beta_{L}\gamma^\mu b^\beta_L,  
\nonumber \\  
{\cal O}_2&=\bar d^\alpha_{L} b^\alpha_R\,  
           \bar d^\beta_{L} b^\beta_R,  
\quad  
{\cal O}_3=\bar d^\alpha_{L} b^\beta_R\,  
             \bar d^\beta_{L} b^\alpha_R,  
\\  
{\cal O}_4&=\bar d^\alpha_{L} b^\alpha_R\,  
           \bar d^\beta_{R} b^\beta_L,  
\quad  
{\cal O}_5=\bar d^\alpha_{L} b^\beta_R\,  
             \bar d^\beta_{R} b^\alpha_L,  
\nonumber  
\end{eqnarray}  
together with three other operators ${\tilde {\cal O}}_{1,2,3}$ 
(and associated coefficients $\tilde C_i$)  
that are chiral conjugates ($L\leftrightarrow R$) of ${\cal O}_{1,2,3}$.  
There are contributions from gluino, neutralino, 
charged Higgs and chargino exchange diagrams \cite{Bertolini}. 
We note that, due to Majorana property of the gluino, 
gluino exchange diagrams can be divided into 
the usual box diagram and the so called crossed diagram. 
By using the double line notation of 't Hooft \cite{tHooft}, 
it is easy to see that the former has a
color $N_c$ factor, while the latter does not.
In general, therefore, the leading SUSY contribution comes from gluino
box diagrams, where we have $\alpha_s^2$ and $N_c$ enhancement,
although it is possible that in some parameter space, 
such as small $\tan\beta$ and when superparticles are light, 
charged Higgs and chargino contributions 
may become important \cite{Branco}.

It is customary to take squarks as   
almost degenerate at scale $\msq$.  
In the following, 
we give contributions from gluino exchange diagrams and make use 
of the mass insertion approximation \cite{Gabbiani,MI}.
In quark mass basis, one defines \cite{MI},  
\begin{equation}
\label{delta}
\delta^{ij}_{qAB} \equiv  
(\msq^2_q)_{AB}^{ij}/{\msq}^2,  
\end{equation}
which is roughly the squark mixing angle,
$\msq^2_q$ are squark mass matrices,  
$A,B=L,R$, and $i,j$ are generation indices.
For notational simplicity, we shall suppress in what follows 
the index pair $ij$ ($13$ for a $\tilde d$--$\tilde b$ mixing angle)
as well as the subscript $d$.

The gluino exchange contributions to the Wilson coefficients 
are \cite{Gabbiani},  
\begin{eqnarray} 
C_1&=&{\alpha^2_s\over \msq^2}\biggl[
{1\over 4}\biggl(1-{1\over N_c}\biggr)^2\,\xgq f_6(\xgq) 
\nonumber \\        
&&\qquad +{1\over 8}\biggl(N_c-{2\over N_c}+{1\over N_c^2}\biggr)
          \tilde f_6(\xgq)\biggr]\,\delta^2_{LL},  
\nonumber \\
C_2&=&{\alpha^2_s\over \msq^2} {1\over 2}
\biggl(N_c-1-{1\over N_c^2}\biggr)\,\xgq f_6(\xgq)\,\delta^2_{LR},  
\nonumber \\ 
\label{Cbox} 
C_3&=&{\alpha^2_s\over \msq^2} {1\over 2}
\biggl(-1+{2\over N_c}\biggr)\,\xgq f_6(\xgq)\,\delta^2_{LR},  
\\  
C_4&=&{\alpha^2_s\over \msq^2}
\biggl\{\biggl(-{1\over2}-{1\over N_c^2}\biggr)\tilde f_6(\xgq)\,\delta_{LR}\delta_{RL}
\nonumber \\
&&\,\, +\biggl[\biggl(N_c-{2\over N_c}\biggr)\,\xgq f_6(\xgq)  
                      -{\tilde f_6(\xgq)\over N_c} \biggr]\,  
                \delta_{LL}\delta_{RR}
\biggr\},  
\nonumber \\  
C_5&=&{\alpha^2_s\over \msq^2}
\biggl\{\biggl({1\over 2 N_c}-{N_c\over 2}\biggr)
\tilde f_6(\xgq)\,\delta_{LR}\delta_{RL}
\nonumber \\
&&\,\,
+\biggl[{\xgq f_6(\xgq)\over N_c^2}  
                      +\biggl({1\over 2}+{1\over 2 N_c^2}\biggr)\tilde f_6(\xgq)\biggr]\,  
                \delta_{LL}\delta_{RR}
\biggr\},  
\nonumber  
\end{eqnarray}  
where $\xgq\equiv \mgl^2/\msq^2$,
and $\tilde C_{1,2,3}$ are given by changing $L\leftrightarrow R$ in   
$C_{1,2,3}$, respectively.
As noted before, the terms containing $N_c$ are from the box diagrams,
while those containing $1$ are from crossed diagrams and 
$1/N_c,1/N_c^2$ are from subleading terms of box and crossed diagrams.
%
%%%%%%%%%%%%%%%%%%%%%%%%%%%%%%%%%%%%%%%%%%%%%%%%%%%%%%%%%%%%%%%%%%%%%%%%%%%%%%%%%%%%%%%%%%
\begin{table}[b!]   
\begin{center}  
\begin{tabular}{c|c||c|c}  
Parameter &Value &Parameter &Value
\\ \hline 
$\bar\rho$ &$0.169$ &$\bar\eta$ &$0.362$
\\ \hline
$f_{B_d}\sqrt{\hat B_{B_d}}$ &$230\pm40$ MeV &$m_B$  &5.2794 GeV
\\ \hline
$\eta_B$ &$0.55$ &$\bar{m}_t(m_t)$  &170 GeV
\\ \hline
$\mu_B$ &2.5 GeV &$m_b(\mu_B)$  &4.88 GeV
\\ \hline
$\alpha_s(\mu_B)$  &0.276 &$\mu_{\rm SUSY}$  &$\sqrt{\widetilde m m_{\tilde g}}$  
\end{tabular} 
\vspace{0.3cm} 
\caption{The input parameters used in this section.
}
\end{center}
\label{table:input}  
\end{table}  
%%%%%%%%%%%%%%%%%%%%%%%%%%%%%%%%%%%%%%%%%%%%%%%%%%%%%%%%%%%%%%%%%%%%%%%%%%%%%%%%%%%%%%%%%
%
%
The loop functions are given by
\begin{eqnarray}
&\big[&-\tilde f_6(\xgq),\,\xgq f_6(\xgq)\big]
\nonumber \\
&&= 
\msq^6{\partial^2\,\over\partial \msq^{\prime2}\partial \msq^2} 
\int_0^\infty dk^2 \biggl\{{k^2 \over(k^2+m_{\tilde g}^2)^2}\times
\label{fi}
\\
\nonumber
&&\qquad\qquad\quad\qquad\qquad{[k^2,\,m^2_{\tilde g}]\over
 (k^2+\msq^2) (k^2+\msq^{\prime2})}\biggr\}
\bigg|_{\msq^{\prime}\to\msq}, 
\end{eqnarray}
which agree with Ref. \cite{Gabbiani}.
Note that $f_6(x_{\tilde g \tilde q})$ is always positive, 
while $\tilde f_6(x_{\tilde g \tilde q})$ is always
negetive.
It is useful to give the asymptotic forms of these functions,
\begin{eqnarray}
&\big[&-\tilde f_6(\xgq),\,\xgq f_6(\xgq)\big]
\nonumber\\
&=&\left\lbrace
\label{asymptoticf}
\begin{array}{ll}
\big[1/({3 \xgq^2}),\,1/\xgq\big],
               &{\rm for}\,\,\, \msq\ll \mgl\,\,\,(\xgq\gg 1),
\\
\big[1/30,\,1/20
\big],
&{\rm for}\,\,\, \mgl=\msq\,\,\,(\xgq\,=\,1),
\\
\big[1/3,\,-{\xgq\ln\xgq}
\big],
&{\rm for}\,\,\, \mgl\ll\msq\,\,\,(\xgq\ll 1).
\end{array}
\right.
\end{eqnarray}  
By using these asymptotic forms, it is easy to show that,
$|\tilde f_6(\xgq)|<|\xgq f_6(\xgq)|$ for $\msq\ll\mgl$ 
and vice versa.
Therefore, one must have a zero in 
$C_1$ ($\tilde C_1$) for some value of $\xgq$ hence a sign change
when passing through it.
One can also show that the cancellation is between the above mentioned
two class of diagrams.
 
After obtaining these Wilson coefficients at SUSY scale $M_{\rm SUSY}$,
we apply renormalization group running to obtain 
$B$ mass scale values.
The renormalization group running of these Wilson coefficients   
including leading order QCD corrections is given in Ref. \cite{Bagger}.
Since $B^0_d$-$\bar B^0_d$ mixing is a $\Delta B=2$ process, 
we need a power of $\delta$ in each of the
internal squark lines to change flavor.
There are altogether six combinations:
$\delta_{LL}^2$, $\delta_{RR}^2$, $\delta_{LL}\delta_{RR}$,
$\delta_{LR}^2$, $\delta_{RL}^2$ 
and $\delta_{LR}\delta_{RL}$, as is evident from Eq. (\ref{Cbox}).  

To obtain $\Delta m_{B_d}$, we use
$\Delta m_{B_d}=2 |M^B_{12}|$, where
\begin{eqnarray} 
M^B_{12} &\equiv& \vert M^B_{12}\vert \, e^{2i\Phi_{B_d}}  
\nonumber \\
 &=& \vert M^{\rm SM}_{12}\vert \, e^{2i\phi_1}  
 + \vert M^{\rm SUSY}_{12}\vert \, e^{i\phi_{\rm SUSY}},  
\label{MB12}
\end{eqnarray}  
and
\begin{eqnarray}
M^{\rm SM}_{12}&=&0.33\, 
              \Biggl({f_{B_d}\sqrt{\hat B_{B_d}}\over 230\,{\rm MeV}}\Biggr)^2
              \Biggl({\bar m_t(m_t)\over 170 {\rm GeV}}\Biggr)^{1.52}
\nonumber \\
&&\times
              \Biggl({\eta_B\over 0.55}\Biggr)
              \Biggl({|V_{td}|\over 8.8\times10^{-3}}\Biggr)^2 e^{2i\phi_1}\,\,
               {\rm ps}^{-1}, 
\end{eqnarray}  
where $M^{\rm SM}_{12}$ is the SM contribution, its value is well  
known \cite{BurasLecture}.
The vacuum insertion matrix elements of ${\cal O}_i$ are given in Ref. \cite{Gabbiani}.
These matrix elements are modified by bag-factors to include non-factorizable effects.
For simplicity, we assume the bag-factors for matrix elements of
${\cal O}_{2-5}$ are equal to $\hat B_{B_d}$, which is calculated for 
${\cal O}_1$.
In the subsequent numerical analysis, 
we take $f_{B_d}\hat B_{B_d}^{1/2}=(230\pm 40)$ MeV \cite{fBd}.
For CKM matrix elements, we take $\vert V_{ub}/\lambda V_{cb}\vert = 0.41$ and  
$\phi_3 =65^\circ$,  
hence $|V_{td}|\times10^{3}=8.0$ to get   
$\Delta m_{B_d}^{\rm SM} \sim 0.54\,{\rm ps}^{-1}$,   
which is close to the experimental value of
$\Delta m_{B_d}=0.484\pm0.010\,{\rm ps}^{-1}$ \cite{LEPBOSC}.
We summarize input parameters used in Table I. %\ref{table:input}. 

\begin{figure}[htb]%t!  
\centerline{{\epsfxsize1.67 in \epsffile{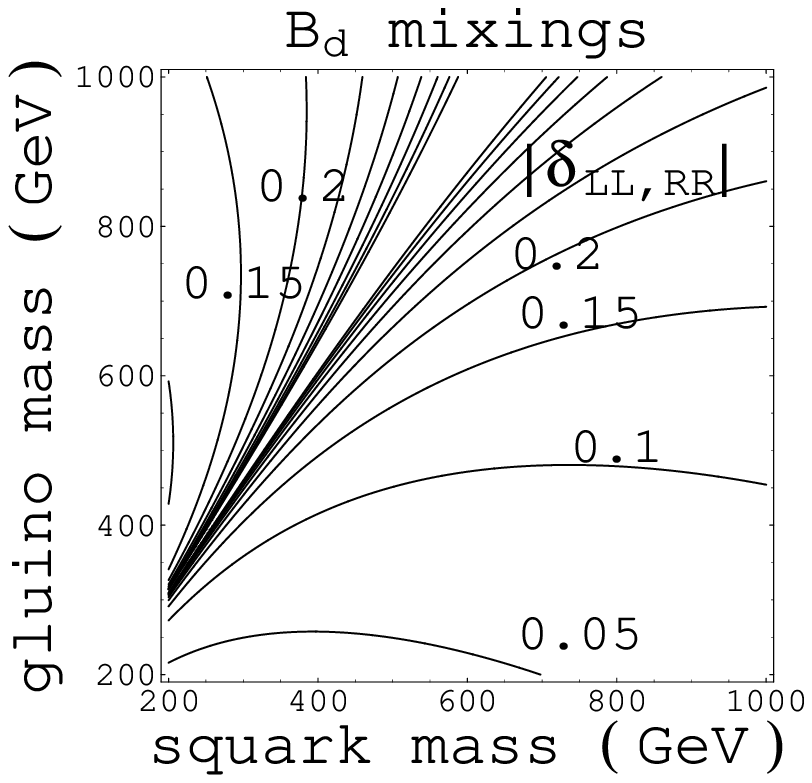}}  
            {\epsfxsize1.67 in \epsffile{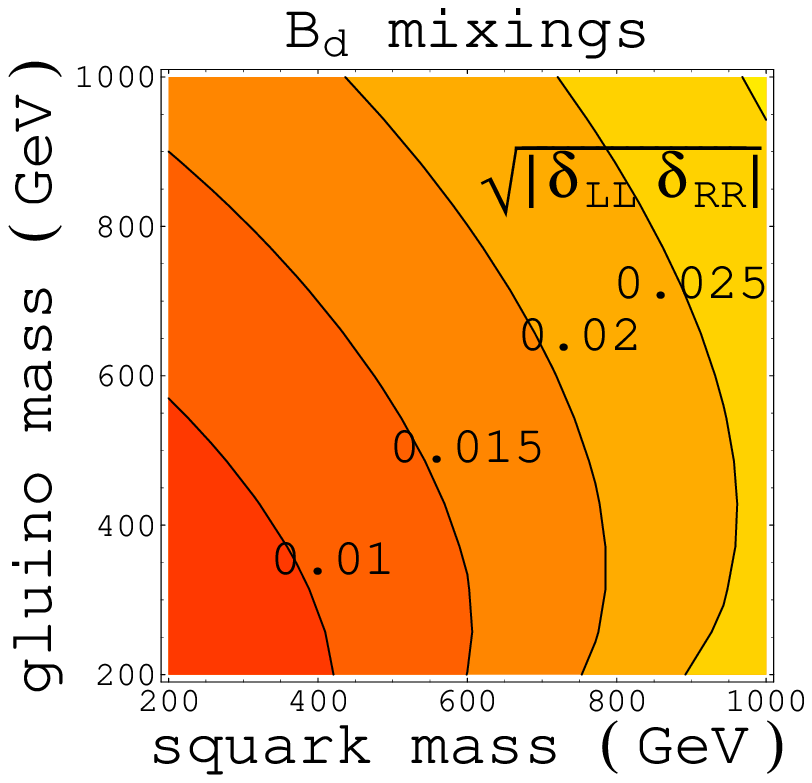}}}  
\smallskip  
\caption{Limits on $\delta_{LL,RR}$ and $\delta_{LL}\delta_{RR}$ 
obtained by assuming $\vert M^{\rm SUSY}_{12}\vert
                    < \vert M^{\rm SM}_{12}\vert$ in $B_d$ mixing.}  
\label{fig:Bdllrr}  
\end{figure}

\begin{figure}[htb]%t!  
\centerline{{\epsfxsize1.67 in \epsffile{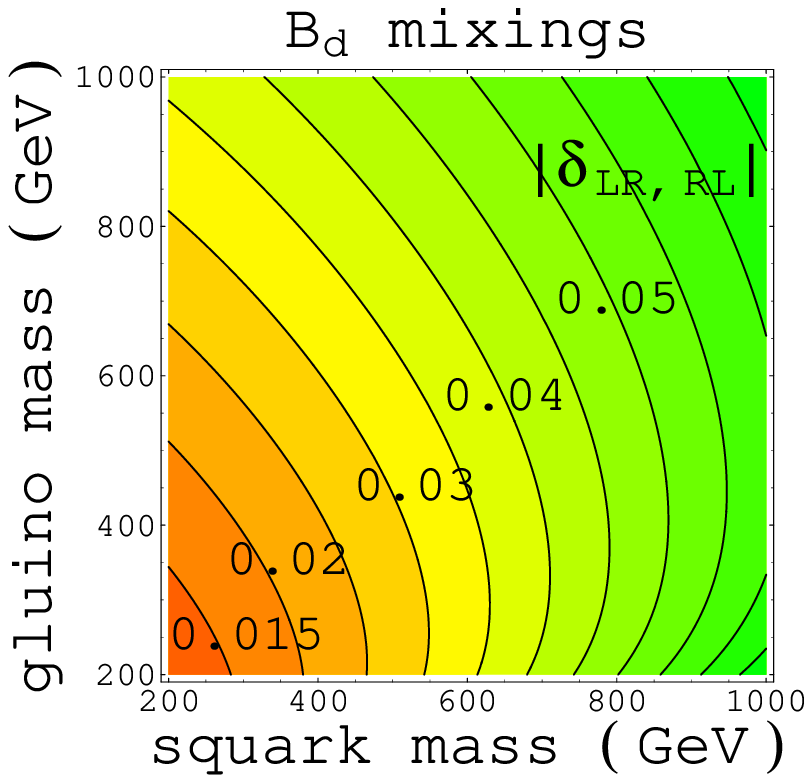}}  
            {\epsfxsize1.67 in \epsffile{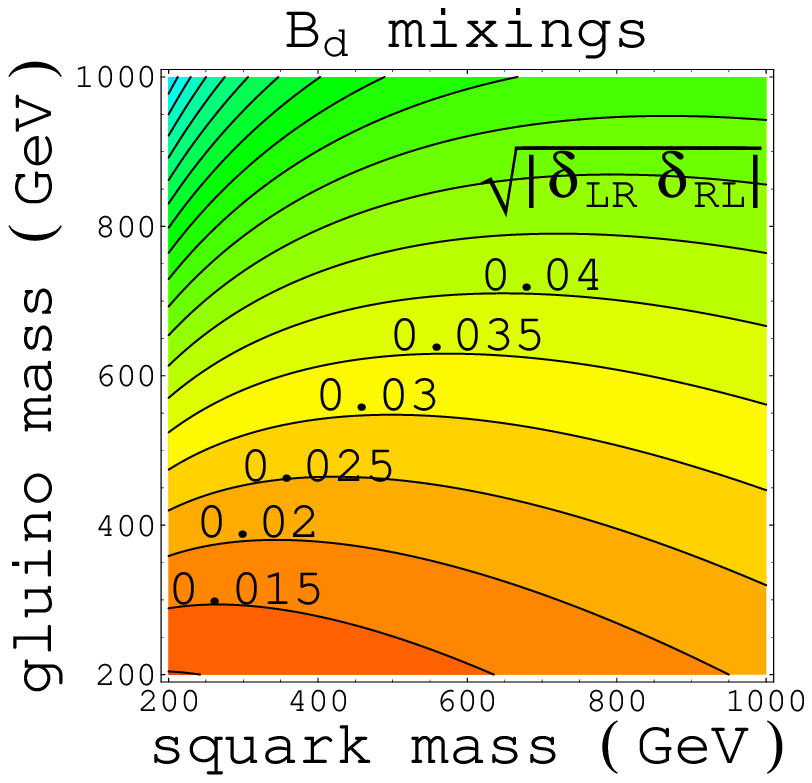}}}  
\smallskip  
\caption{Limits on $\delta_{LR,RL}$ and $\delta_{LR}\delta_{RL}$ 
obtained by assuming $\vert M^{\rm SUSY}_{12}\vert
                    < \vert M^{\rm SM}_{12}\vert$ in $B_d$ mixing.}  
\label{fig:Bdlrlr}  
\end{figure}

In Figs. \ref{fig:Bdllrr} and \ref{fig:Bdlrlr}, we show 
estimated limits of all six $\sqrt{|\delta\,\delta|}$s over the parameter
space of sub-TeV gluino and squarks.
The limits are taken such that \cite{ChuaHou} the SUSY contribution 
to $B_d$ mixing matrix element, $|M_{12}|$, 
is comparable with the SM result,
hence large interference effects could in principle occur 
that can give low $a_{J/\psi K_S}$.
Note that by assuming the same bag-factor for ${\cal O}_i$, the uncertainty
on $f_{B_d}\hat B_{B_d}^{1/2}$ does not show up in these figures.
Squark mixing angles that are much larger than those shown 
would give too large a contribution to $B_d$ mixing 
and would require fine tuning to satisfy the experimental result.
For mixing angles that are much smaller than those shown,
they will not be able to generate large enough interference effect 
to reduce the asymmetry.  
Therefore, the limits shown in these figures may serve as upper limits from
$\Delta m_{B_d}$ constraint on one hand, and serve as roughly the required 
values to give impact on $a_{J/\psi K_S}$.

%%%%%%%%%%%%%%%%%%%%%%%%%%%%%%%%%%%%%%%%%%%%%%%%%%%%%%%%%%%%%%%%%%%%%%%%%%%%
%%%%%%%
%Fig 1% 
%%%%%%%
%%%%%%%
%Fig 2% 
%%%%%%%
%%%%%%%%%%%%%%%%%%%%%%%%%%%%%%%%%%%%%%%%%%%%%%%%%%%%%%%%%%%%%%%%%%%%%%%%%%%%

From Figs. \ref{fig:Bdllrr} and \ref{fig:Bdlrlr} we see that
the limits on $\sqrt{|\delta_{LL}\delta_{RR}|}$, $|\delta_{LR,RL}|$ and
$\sqrt{|\delta_{LR}\delta_{RL}|}$ are all of order few \%,
with $\sqrt{|\delta_{LL}\delta_{RR}|}$ as
the most sensitive source for $B_d$ mixing.
This can be understood from Eq.~(\ref{Cbox}), 
where $\sqrt{|\delta_{LL}\delta_{RR}|}$ in $C_4$ has 
the largest $N_c$ factor,
while there is also RG enhancement \cite{Bagger}. 
Furthermore, we see from Eq.~(\ref{Cbox}) that 
the dominant $\delta_{LL}\delta_{RR}$ and $\delta^2_{LR,RL}$ terms 
are proportional to $\xgq f_6(\xgq)/\msq^2$,
while dominant $\delta_{LR}\delta_{RL}$ terms
is proportional to $\tilde f_6(\xgq)/\msq^2$.
Therefore, the bounds on $\sqrt{|\delta_{LL}\delta_{RR}|}$ 
and $|\delta_{LR,RL}|$ are roughly proportional to 
$\sqrt{\msq^2/[\xgq f_6(\xgq)]}$,
while the bound on $\sqrt{|\delta_{LR}\delta_{RL}|}$ is roughly proportional
to $\sqrt{\msq^2/|\tilde f_6(\xgq)|}$,
such that Figs.~\ref{fig:Bdllrr}(b) and \ref{fig:Bdlrlr}(a) show
similar behavior that is different from Fig. \ref{fig:Bdlrlr}(b).

The order of magnitude of these figures can be understood 
by a simple dimensional analysis.
Comparing the SUSY vs SM box diagram contributions, we find 
$\delta \lesssim V_{tb} V^*_{td} ({\alpha_{\rm W}/\alpha_S})
({m_t\msq/M^2_{\rm W}}) ({1/\sqrt N_c}) 
        \lesssim 2.8 \times 10^{-2} \, ({\msq/500\,{\rm GeV}})$,  
which is very close to the limits on $|\delta_{LR,RL}|$,
$\sqrt{|\delta_{LR}\delta_{RL}|}$ and $\sqrt{|\delta_{LL}\delta_{RR}|}$
from Figs.~\ref{fig:Bdllrr}(b) and \ref{fig:Bdlrlr}.
However, the limit on $|\delta_{LL,RR}|$ as shown in 
Fig. \ref{fig:Bdllrr}(a) are of order few 10\% 
and do not obey this estimation. 
This rather different behavior is because of the possible cancellation
between $\xgq f_6(\xgq)$ and $\tilde f_6(\xgq)$ in $C_1(\tilde C_1)$,
which can weaken the bounds. 
A total cancellation is reflected in the
valley along $x_{\tilde g,\tilde q}\sim2.43$ where  
$|\delta_{LL,RR}|$ is not constrained by 
$|M^{\rm SUSY}_{12}|\sim |M^{\rm SM}_{12}|$.
From Eqs. (\ref{Cbox}) and (\ref{asymptoticf}), 
we see that $C_1$ is dominated by $\xgq f_6(\xgq)$ for $\xgq\gg 1$ 
and dominated by $\tilde f_6(\xgq)$ for $\xgq\ll 1$.
Therefore, appart from 
the distortion due to the cancellation discussed earlier, 
the upper left part of Fig. \ref{fig:Bdllrr}(a) is similar to  
Fig. \ref{fig:Bdllrr}(b) and Fig. \ref{fig:Bdlrlr}(a), 
while the lower right part
is similar to Fig. \ref{fig:Bdlrlr}(b).
Note that whenever we obtain a bound that is greater than $O(1)$ in 
the squark mixing angle $\delta$,
it should be interpreted as signaling 
the need of a large squark mass splitting
which invalidates the approximation of Eq. (\ref{delta}).  

It is clear that, to obtain low $a_{J/\psi K_S}$, 
we need suitable SUSY phase to
have destructive interference with SM.
But the minimum requirement is the SUSY amplitude should be large enough to 
allow for such an interference effect.
As shown in this section, it is possible for SUSY models to give 
$B_d$ mixing that is comparable to SM result hence lead to
large interference effect.
Mixing angles of few \%  in left-right squark mixings or
few \% to few 10\% in left-left, right-right squark mixings
are sufficient to achieve this.
The case with both left-left and right-right mixings
 ($\delta_{LL}\delta_{RR}$) 
is most sensitive to mixing angles, 
while left-left or right-right mixing alone are the least sensitive,
and could even be totally insensitive for a fine-tuned parameter space
near $\xgq\equiv\mgl^2/\msq^2\sim2.43$.

%%%%%%%%%%%%%%%%%%%%%%%%%%%%%%%%%%%%%%%%%%%%%%%%%%%%%%%%%%%%%%%%%%%%%%%%%%%
%Fig 3% 
%%%%%%%
\begin{figure}[t!]%t!  
\centerline{{\epsfxsize1.7 in \epsffile{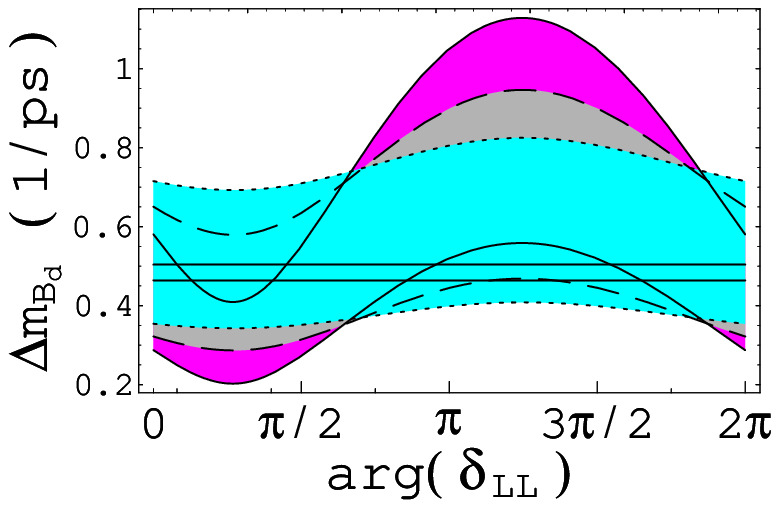}}  
            {\epsfxsize1.7 in \epsffile{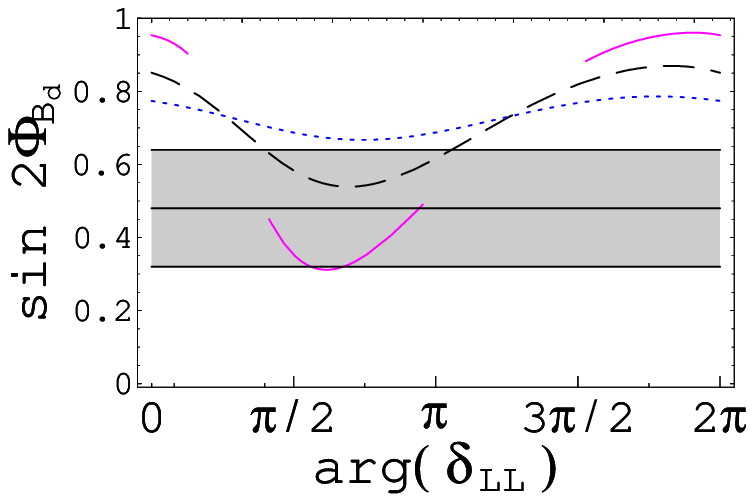}}}  
\smallskip  
\caption{ Dotted, dashed and solid lines correspond to 
          (a) $\Delta m_{B_d}$, (b) $\sin 2\Phi_{B_d}$, induced by 
          $\delta_{RR}=|\delta_{LL}|$=(0.3, 0.5, 0.7)$\times0.013$ with 
          $\msq,\,\mgl=500, 400$ GeV, respectively.
          The horizontal band in the left (right) figure is 
          2$\sigma$(1$\sigma$) experimental range.}  
\label{fig:Bdmixing}  
\end{figure} 
%%%%%%%%%%%%%%%%%%%%%%%%%%%%%%%%%%%%%%%%%%%%%%%%%%%%%%%%%%%%%%%%%%%%%%%%%%%%

For illustration, we pick a point from Figs. \ref{fig:Bdllrr}, \ref{fig:Bdlrlr},
say $(\msq,\,\mgl)=(500,\,400)$ GeV, and show the SUSY effects on $B_d$ mixing.
For the LL-RR mixing case, the bound is 0.013, as shown in
Fig. \ref{fig:Bdllrr}(b).
We show in Fig. \ref{fig:Bdmixing},
$\Delta m_{B_d}$ and $\sin 2\Phi_{B_d}$, induced by 
$\delta_{RR}$ = $|\,\delta_{LL}|$ = (30\%, 50\%, 70\%)$\times0.013$, 
vs. arg($\delta_{LL}$), respectively.
The $\sqrt{|\,\delta_{LL}\delta_{RR}|}$ are (30\%, 50\%, 70\%) of the
bound shown in Fig. \ref{fig:Bdllrr}(b) for the particular $\msq,\,\mgl$.
Larger oscillating amplitudes in the figures correspond to larger
$\delta_{LL,RR}$.
Similar results will be obtained by using (30\%, 50\%, 70\%) of the 
corresponding bounds of other points on $(\msq,\,\mgl)$ plane,
shown in Fig. \ref{fig:Bdllrr} and Fig. \ref{fig:Bdlrlr}. 
The horizontal band in the left (right) figure is 2$\sigma$(1$\sigma$) 
experimental range.
The uncertainty of the predicted $\Delta m_{B_d}$ is due to
the $\sim17\%$ uncertainty of $f_{B_d}\hat B_d^{1/2}$ as shown in 
Table I. %\ref{table:input}.
This factor does not enter arg$(M^B_{12})$, 
and thus $\sin 2\Phi_{B_d}$ shown in Fig. \ref{fig:Bdmixing} (b), 
as we assume all bag-factors for different ${\cal O}_i$ are the same.
By using a $\delta$ at 30\% of the bound value, 
$B_d$ mixing does not differ much from the SM prediction,
while for the 50\% case, it starts to show interesting deviation with 
$\sin 2\Phi_{B_d}$ as low as 0.53.
Note that the SM gives $\sin2\phi_1=0.73$ by using our input parameters. 
For a larger $\delta$, such as 70\% of the bound, 
it can further lower $\sin 2\Phi_{B_d}$ to 0.3.
Although, in this case not all arg($\delta_{LL}$) are allowed,
due to the $\Delta m_{B_d}$ constraint,
we still have plenty of allowed region for this phase.  
From these figures, we see that by using 50\%--70\% of the $\delta$ bounds,
low $\sin 2\Phi_{B_d}$ can be easily obtained.

\section{$B\to\pi\pi$ and $\rho\gamma$ Decays in SUSY Models}

%%%%%%%%%%%%%%%%%%%%%%%%%%%%%%%%%%%%%%%%%%%%%%%%%%%%%%%%%%%%%%%%%%%%%%%%%%%%%%%%
\begin{table}[b!]   
\label{table:input1}  
\begin{center}  
\begin{tabular}{c|c||c|c}  
Parameter &Value &Parameter &Value
\\ \hline
$F_0(0)$ &$0.30\pm0.04$ &$f_\pi$  &$133$ MeV
\\ \hline
$m_\pi$  &140 MeV &$\tau_B$  &1.548 ps
\\ \hline
$m_u(\mu_B)$ &$2$ MeV &$m_d(\mu_B)$ &$4$ MeV
\\ \hline
$m^{\rm const.}_{u,d}$ &0.2 GeV &$m^{\rm const.}_s$ &0.5 GeV
\\ \hline
$m^{\rm const.}_c$ &1.5 GeV &$\langle q^2\rangle$ &$m^2_b/3$
\end{tabular} 
\vspace{0.3cm} 
\caption{The input parameters used in this section.
}
\end{center}  
\end{table}  
%%%%%%%%%%%%%%%%%%%%%%%%%%%%%%%%%%%%%%%%%%%%%%%%%%%%%%%%%%%%%%%%%%%%%%%%%%%%%

The effective Hamiltonian for charmless $b\to d$ decays is,
\begin{eqnarray}
H_{\rm eff}&=&{4G_F\over \sqrt 2} \biggl[V_{ub} V^*_{ud} (c_1 O_1+c_2 O_2)
        -V_{tb} V^*_{td} \sum^{10}_{i=3} c_i O_i
\nonumber \\
        &&\qquad\qquad-V_{tb} V^*_{td} (C_{g} \widetilde O_{g}
                         +C^\prime_{g} \widetilde O^\prime_{g})\biggr],
\end{eqnarray}
where, as a matter of convention,
we factor out a CKM factor $V_{tb} V^*_{td}$ even for SUSY contributions.
The operators are defined as,
\begin{eqnarray}
O_1&=&\bar u \gamma_\mu L b\,\bar d \gamma^\mu L u,
\nonumber \\
O_2&=&\bar u_\alpha \gamma_\mu L b_\beta\,\bar d_\beta \gamma^\mu L u_\alpha,
\nonumber\\
O_{3(5)}&=&\bar d \gamma_\mu L b\,\bar q \gamma^\mu L(R) q,
\nonumber \\
O_{4(6)}&=&\bar d_\alpha\gamma_\mu L b_\beta\,\bar q_\beta\gamma^\mu L(R)q_\alpha,
\\
O_{7(9)}&=&{3\over2}
           \bar d \gamma_\mu L b\,Q_q\bar q \gamma^\mu R(L) q,
\nonumber \\
O_{8(10)}&=&{3\over2}
\bar d_\alpha\gamma_\mu L b_\beta\,Q_q\bar q_\beta\gamma^\mu R(L)q_\alpha,
\nonumber\\
\widetilde O^{(\prime)}_{g}&=&{\alpha_s\over4\pi} \bar d i\sigma_{\mu\nu} T^a
                  {2m_b q^\nu\over q^2} R(L) b\,\,\bar q\gamma^\mu T^a q,
\nonumber
\end{eqnarray}
where $L,R=(1\mp\gamma_5)/2$, 
$\widetilde O^{(\prime)}_{g}$ arises from the dimension 5 color dipole
operator, and $q=p_b-p_d$.
Note that with New Physics, one may also have chiral conjugates of $O_{1-10}$ with
Wilson coefficients defined as $c^\prime_{1-10}$.
The Wilson coefficient $C^{(\prime)}_{g}$ and color dipole operators are defined in
the effective Hamiltonian     
for $b\rightarrow d\gamma,\,dg$ transitions,    
\begin{eqnarray}  
H_{{\rm eff}} &=&-{V_{tb}V_{td}^{*}m_{b}G_{F}\over4\sqrt{2}\pi^2}
\big\{
e\,\bar{d}\left[ C_{\gamma}\,R+  
C_{\gamma}^{\prime}\,L\right] \sigma _{\mu \nu }F^{\mu \nu }b  
\nonumber \\  
&&\qquad+g\,\bar{d}\left[ C_{g}\,R+  
C_{g}^{\prime }\,L\right]  
\sigma _{\mu \nu }T^{a}G_{a}^{\mu \nu }b\big\},  
\end{eqnarray}  
where we have neglected $m_{d}$,   
$C_{\gamma,g}=C_{\gamma,g}^{{\rm SM}}+C_{\gamma,g}^{{\rm New}}$   
are the sum of SM and new physics contributions,   
while $C_{\gamma,g}^{\prime}$ come purely from New Physics.

For $\bar B^0\to\pi^+\pi^-$, using factorization approach, we find 
\cite{Ali,Deshpande,He}
\begin{eqnarray}
iM&=&-i{G_F\over\sqrt 2} f_\pi F_0^{B\to\pi}(m^2_\pi)(m_B^2-m_\pi^2)
\biggl\{V_{ub} V^*_{ud} a_1
\nonumber\\
&&-V_{tb} V^*_{td} \biggl[\Delta a_4+\Delta a_{10}+(\Delta a_6+\Delta a_8) R
\nonumber \\
&&\qquad\qquad\qquad+{\alpha_s\over4\pi}{2m_b^2\over q^2} \widetilde S_{\pi\pi} 
 (C_{g}-C^\prime_{g})
\biggr]
\biggr\},
\label{amplitude}
\\
R&=&{2 m_\pi^2\over (m_b-m_u)(m_u+m_d)},
\\
\widetilde S_{\pi\pi}&=&-{N_c^2-1\over 2 N^2_c}
\biggl\{(1+R)-R{m_B^2(m_b-m_u)\over2 m_b (m_B^2-m_\pi^2)}\times
\nonumber \\
&&\qquad\biggl({3\over2}{f^+(m^2_\pi)\over F_0(m^2_\pi)}
+{f^-(m^2_\pi)\over 2 F_0(m^2_\pi)}\biggr) 
+{m_B^2\over2m_b(m_b-m_u)}
\nonumber\\
\label{Spipi}
&&\qquad\quad
-{4m_b \, h(m^2_\pi)\over F_0(m^2_\pi)}
{m_B^2(m_B^2-4m_\pi^2)\over8m_b^2(m_B^2-m_\pi^2)}
\biggr\},
\end{eqnarray}
where $\Delta a_i$ is defined as $a_i-a^\prime_i$ with 
$a^{(\prime)}_i\equiv c^{(\prime)}_i+c^{(\prime)}_{i\mp1}/N_c$, 
for even (odd) $i$. 
Using input parameters shown in Table II,
the chiral factor $R=1.33$.
The expression for $\widetilde S_{\pi\pi}$ is 
somewhat different from the one given in Ref. \cite{Deshpande} 
because of the treatment of $q^\nu$ in $\widetilde O^{(\prime)}_{g}$.
We have used 
\begin{eqnarray}
\bar d i\sigma_{\mu\nu}&q^\nu& T^a R b\,\,\bar q\gamma^\mu T^a q
\nonumber \\
&=& -(\bar d \gamma_\mu \pbslash T^a R b
      +\bar d \pdslash \gamma_\mu T^a R b)\,(\bar q\gamma^\mu T^a q)
\nonumber \\
&&+\bar d T^a R b\,p_{b\mu} [\bar q\,(2\pbslash-\qslash)\gamma^\mu T^a q].
\end{eqnarray}
%
%
%
%%%%%%%%%%%%%%%%%%%%%%%%%%%%%%%%%%%%%%%%%%%%%%%%%%%%%%%%%%%%%%%%%%%%%%%%%%%%%%%  
%\onecolumn  
%%%%%%%%%%%%%%%%%%%%%%%%%%%%%%%%%%%%%%%%%%%%%%%%%%%%%%%%%%%%%%%%%%%%%%%%%%%%
\begin{table}[t!] %htb]  
\label{table:ai}  
\twocolumn[\hsize\textwidth\columnwidth\hsize\csname
@twocolumnfalse\endcsname
\begin{center}  
\begin{tabular}{c||l|l||c||l|l}  
$a_i$ & $\phantom{-}b\to d $ & $\phantom{-}\bar b\to \bar d$
 & $a_i$ & $\phantom{-}b\to d$ & $\phantom{-}\bar b\to \bar d$
\\ \hline 
$a_1$ & $\phantom{-}1.0463$ & $\phantom{-}1.0463$ &
 $a_6$ & $-0.0473+0.0091i$ & $-0.0674+0.0029i$
\\ \hline
$a_2$ & $\phantom{-}0.0502$ & $\phantom{-}0.0502$ &
 $a_7$ & $\phantom{-}0.0002$ & $-0.0000$
\\ \hline
$a_3$ & $\phantom{-}0.0049$ & $\phantom{-}0.0049$  &
 $a_8$ & $\phantom{-}0.0005$ & $\phantom{-}0.0004$
\\ \hline
$a_4$ & $-0.0337+0.0091i$ & $-0.0538+0.0029i$ &
 $a_9$ & $-0.0093$ & $-0.0095$
\\ \hline
$a_5$ & $-0.0044$ & $-0.0044$ &
 $a_{10}$ & $-0.0013$ &$-0.0014$
%\\ \hline  
\end{tabular} 
\vspace{0.3cm} 
\caption{The $a_i$s in SM for $b\to d$ and $\bar b\to \bar d$ for
$\phi_3=65^\circ$.
}
\end{center}  
] 
\end{table}  

\noindent where the $\not\!\!{{q\,\,}}$ 
term can be dropped because of current conservation.
By using heavy quark symmetry, $p_b\to p_B$,  
it is then straightforward to use factorization approach to obtain 
$\widetilde S_{\pi\pi}$ given in Eq. (\ref{Spipi}).
For the form factors involved, 
we have the relations \cite{Deshpande} 
$f^+(m^2_\pi)=F_1(m^2_\pi)$, 
$f^-(m^2_\pi)=(m_B^2/m^2_\pi-1) \,[F_0(m^2_\pi)-F_1(m^2_\pi)]$
and $4m_b \, h(m^2_\pi)=f_+(m^2_\pi)-f^-(m^2_\pi)$.
Using $F_0(0)=F_1(0)=0.30\pm0.04$ and monopole form factors
for $F_{0,1}$, with pole masses given in Ref. \cite{Ali}, 
it is easy to show that $f^+(m^2_\pi)/F_0(m^2_\pi)\sim1$, 
$f^-(m^2_\pi)/F_0(m^2_\pi)\sim 0$, and
$4m_b \, h(m^2_\pi)/F_0(m^2_\pi)\sim1$.
Therefore,
\begin{equation}
\label{Spipiapprox}
\widetilde S_{\pi\pi}\simeq-{N_c^2-1\over 2 N^2_c}\biggl({11\over 8}+{R\over 4}\biggr)
                     \simeq-0.76,
\end{equation}
which is not far from the value of $-0.80$ computed from
Eq.~(\ref{Spipi}),  and also close to the value given in Ref. \cite{He}.
Note that it is insensitive to $N_c$ and
the chiral factor $R$.
The opposite sign between $(a_i,\,C_{g})$ and $(a^\prime_i,\,C^\prime_{g})$ 
can be easily understood by using parity transformation.
We note that the color dipole term is sensitive to
$\langle q^2\rangle$, which is usually taken to be 
between $m^2_b/4$ and $m^2_b/2$.
We use $\langle q^2\rangle\sim m^2_b/3$. 
\footnote{When compare to the QCD factorization approach, 
we may use effectively $\langle q^2\rangle\sim m^2_b/3$ -- $m^2_b/2.4$ 
in the color dipole contribution \cite{KC}.}
The $q^2$ dependence will affect the color dipole contribution
by $\pm 33\%$ at amplitude level.

Using input parameters shown in Tables \ref{table:input}, \ref{table:input1}
and following the approach in Ref. \cite{Ali}, we obtained
the numerical values of $a_i$s in SM, as shown in Table \ref{table:ai}. 
In SM, $C^{\rm SM}_{g}=-0.15$ while $C^{\prime\rm SM}_{g}$ is highly 
suppressed by the V$-$A nature of weak interaction 
and the smallness of $m_d$.
The $C^{\rm SM}_{g}$ contribution is about 3\% of tree amplitude.
Under factorization approximation, 
the SM gives Br$(\bar B^0\to\pi^+\pi^-)\sim10\times10^{-6}$, 
%and Br$(\bar B^0\to\pi^+\pi^-)=10.2\times10^{-6}$ 
with $\sim3.3$\% asymmetry,
as defined as,
\begin{equation}
a_{\pi^+\pi^-}={{\rm Br}(B^0\to\pi^+\pi^-)-{\rm Br}(\bar B^0\to\pi^+\pi^-)
                      \over
                      {\rm Br}(B^0\to\pi^+\pi^-)+{\rm Br}(\bar B^0\to\pi^+\pi^-)}.
\end{equation}
Since this process is tree dominant, we need large contribution if
the rate is to be reduced by New Physics contribution.

In SUSY models, we can have gluino, neutralino,
chargino and charged Higgs exchange contributions to $B\to\pi\pi$. 
Because of $N_c$ enhancement and 
different sensitivities of photonic vs gluonic penguins \cite{Kagan},
and since one does not suffer from Br$(B\to X_s\gamma)$ constraint,
we see that gluino exchange gives dominant and interesting contribution
to $B\to\pi\pi$ compared to other superparticles.
There are two types of diagrams:
The gluino box and the gluino penguin.
The former as well as the $F_1$ term
 (the quark chirality conserving vertex term) 
of the latter contribute to $a_i$,
and only depend on one power of $\delta_{LL,RR}$.
The $F_2$ term (the quark chirality flipped vertex term)
of the gluino penguin contributes through
$C^{(\prime)}_{g}$ with all types of squark mixings.
We use similar formulas as in Ref. \cite{Gabbiani} for SUSY contribution
and those of Ref. \cite{BurasLecture,c7c8mix} for RG running.

The gluino box and $F_1$ gluino penguin give,
\begin{eqnarray}
c_3(M_{\rm SUSY})&=&{\alpha_s^2\delta_{LL}\over2\sqrt2 G_f V_{tb}V^*_{td}\msq^2}
\biggl[-{1\over N_c^2}B_1(\xgq)-{1 \over 2}\times
\nonumber \\
&&\qquad\quad\biggl(1+{1\over N_c^2}\biggr)B_2(\xgq)
-{1\over N_c}P(\xgq)\biggr],
\nonumber\\
c_4(M_{\rm SUSY})&=&{\alpha_s^2\delta_{LL}\over2\sqrt2 G_f V_{tb}V^*_{td}\msq^2}
\biggl[\biggl(-N_c+{2\over N_c}\biggr)B_1(\xgq)
\nonumber \\
&&\label{ci}
\qquad\qquad+{1\over N_c}B_2(\xgq)+P(\xgq)\biggr],
\\
c_5(M_{\rm SUSY})&=&{\alpha_s^2\delta_{LL}\over2\sqrt2 G_f V_{tb}V^*_{td}\msq^2}
\biggl[\biggl(1+{1\over N_c^2}\biggr)B_1(\xgq)
\nonumber \\
&&\qquad\qquad+{1\over 2N_c^2}B_2(\xgq)-{1\over N_c}P(\xgq)\biggr],
\nonumber \\
c_6(M_{\rm SUSY})&=&{\alpha_s^2\delta_{LL}\over2\sqrt2 G_f V_{tb}V^*_{td}\msq^2}
\biggl[-{2\over N_c}B_1(\xgq)+{1 \over 2}\times
\nonumber \\
&&\qquad\qquad\biggl(N_c-{2\over N_c}\biggr)B_2(\xgq)+P(\xgq)\biggr],
\nonumber
\end{eqnarray}
where $c^\prime_i$ are obtained from $c_i$ 
by replacing $L\leftrightarrow R$, $B_i(x)$ are from gluino box, 
$P(x)\equiv [C_2(G)/2-C_2(R)]\,P_1(x)+C_2(G) P_2(x)/2$
are from the $F_1$-term, and $C_{2}(G)=N_c$ and
$C_{2}(R)=(N^{2}_c-1)/2N_c$ are Casimirs. 
The $P_{1(2)}(\xgq)$ term
corresponds to gluon attached to squark (gluino) line. 
The gluino box contribution to $c_{3,4\,(5,6)}$ are due to 
$\tilde d_L$--$\tilde b_L$ mixings in one of the squark line, 
while the other squark line is $\tilde q_{L(R)}$. 
The leading terms in $N_c$ are the $B_1(x)$ and
$B_2(x)$ terms of $c_4$ and $c_6$, respectively.
Explicit forms of $B_i(x),\, P_i(x)$ can be found in Ref. \cite{Gabbiani}.  
They can be expressed as,
\begin{eqnarray}
&&\big[-4B_1(\xgq),\,B_2(\xgq)\big]
\nonumber \\
&&=
\msq^4{\partial\,\over\partial \msq^{\prime2}}
\int_0^\infty dk^2 \biggl\{{k^2 \over(k^2+m_{\tilde
g}^2)^2(k^2+\msq^{\prime2})}\times
\label{Bi}
\\
\nonumber
&&\qquad\qquad\quad\qquad\qquad\qquad\qquad{[k^2,\,m^2_{\tilde g}]\over
  (k^2+\msq^{2})}\biggr\}
\bigg|_{\msq^{\prime}\to\msq},
\label{Pi}
\\
&&6\,\big[P_1(\xgq),\,-P_2(\xgq)\big]
\nonumber \\
&&=\msq^4{\partial\,\over\partial \msq^{2}}
\int_0^\infty dk^2 { k^4\over (k^2+m^2_{\tilde g})(k^2+\msq^2)}\times
\nonumber \\
&&\qquad\qquad\quad\qquad\qquad
\bigg[{k^2 \over (k^2+\msq^2)^3},\,{2k^2+3\mgl^2 \over(k^2+m^2_{\tilde
g})^3}\bigg].
\end{eqnarray}
Note that $B_1(x)$, $P_2(x)$ are always positive, 
while $B_2(x)$, $P_1(x)$ are always negative.
It is useful to give the asymptotic form of these functions,
\begin{eqnarray}
\label{asymptoticBP}
&\big[&B_1(x),\,B_2(x),\,P(x)\big]
\nonumber \\
&&={1\over4}\left\lbrace
\begin{array}{ll}
\big[\ln x/x^2,\,-2/x,\,-1/(9x)\big],
               &\,\,x\gg1,
\\
\big[1/12,\,-1/3,\,8/45\big],
               &\,\,x\,=\,1,
\\
\big[1/2,\,4x\ln x,\,-2\ln x \big],
&\,\,x\ll 1.
\end{array}\right.
\label{BBP}
\end{eqnarray}

The gluonic and photonic penguins are closely related.
The formulas for $C^{(\prime)}_{\gamma}$ and $C^{(\prime)}_{g}$
from gluino exchange are,
\begin{eqnarray}  
C_{\gamma} &=&   
\frac{\pi \alpha _{s}}{\sqrt{2}{G}_{F}V_{tb}V_{td}^{*}}  
 \frac{Q_{d}\,2C_{2}(R)}{{\msq}^2}  
 \biggl\{ \delta_{LL}\, g_{2}(x_{\tilde g \tilde q})  
\nonumber \\
\label{c7New}         
&&\qquad\qquad\qquad\qquad\qquad-\frac{\mgl}{m_{b}}\,  
         \delta_{LR}\, g_{4}(x_{\tilde g \tilde q})  
 \biggr\},
\\  
C_{g} &=&  
 \frac{\pi \alpha_{s}}{\sqrt{2}{G}_{F} {\msq}^2 V_{tb}V_{td}^{*}}  
 \biggl\{ \delta_{LL}   
 \biggl([2C_2(R)-C_2(G)]\, g_2(x_{\tilde g \tilde q})
\nonumber \\  
&&\qquad\qquad\qquad\qquad\qquad -C_2(G)\, g_1(x_{\tilde g \tilde q})   
  \biggr)  
\nonumber \\  
  &&\quad\qquad\qquad+\frac{\mgl}{m_b} \,  
    \delta_{LR}  
  \biggl([C_2(G)-2C_2(R)]\, g_4(x_{\tilde g \tilde q})
\nonumber \\
&&\qquad\qquad\qquad\qquad\qquad+C_2(G)\, g_3(x_{\tilde g \tilde q})  
  \biggr) \biggr\},   
\label{c8New}  
\end{eqnarray}  
with the chirality partners $C_{\gamma,g}^{\prime }$
obtainable by interchanging $L\leftrightarrow R$ in the $\delta$'s,
$Q_{d}$ is the electric charge of the down type quarks and 
the functions $g_i(\xgq)=-\msq^4$ $({\partial/\partial \msq^2})$
$[F_i(\xgq)/\msq^2]$, where $F_i(x)$ are given in Ref. \cite{Bertolini}, 
and can be expressed in terms of loop integrals,
\begin{eqnarray}
&\big[F_1&(\xgq),\,F_2(\xgq),\,F_2(\xgq),\,F_3(\xgq),F_4(\xgq)\big]
\nonumber \\
&=&\msq^2\int_0^\infty dk^2 { k^2\over (k^2+m^2_{\tilde g})(k^2+\msq^2)} 
\bigg[{k^2 m^2_{\tilde g}\over 2(k^2+m^2_{\tilde g})^3},
\nonumber \\
&&\qquad\,{k^2\msq^2\over 2(k^2+\msq^2)^3},
\,{k^2\over(k^2+m^2_{\tilde g})^2},
\,{\msq^2\over(k^2+\msq^2)^2}\bigg].
\label{Fi}
\end{eqnarray}
For later purpose, we give the asymptotic behavior of $g_i(\xgq)$,
\begin{eqnarray}
&-& 6\big[g_1(x),\,g_2(x),\,g_3(x),g_4(x)\big]
\nonumber \\
&=&
\left\{
\begin{array}{ll}
%\hspace{-1mm}
\big[1,\, 3\ln x,\, 3,\, 6\ln x\big]/x^2,\,\,\,\
& {\rm for}\,\,\,\, x\gg1,
\\
%\hspace{-1mm}
\big[1/10,\,3/20,\,1/2,\,1/2\big],\,\,\,
& {\rm for}\,\,\,\, x\,=\,1,
\\
%\hspace{-1mm}
\big[1,\,1/2,\,-6\ln x,\,1\big],
& {\rm for}\,\,\,\, x\ll1.
\end{array}
%\hspace{-3mm}
\right.
\label{asymptoticg}
\end{eqnarray}  
From Eqs. (\ref{c7New}--\ref{Fi}), it is clear that 
$g_{2,4}(\xgq)$ correspond to photon and gluon 
attached to the internal squark line, 
while $g_{1,3}(\xgq)$ correspond to the opposite case 
but only with the gluon attachment.
Note that in the large $N_c$ limit, $C_{2}(G),\,2C_{2}(R)\to N_c$, 
while $[2C_{2}(R)-C_{2}(G)]\to {\cal O}(1/N_c)$ is suppressed.
One always has an $N_c$ factor in $C^{(\prime)}_{\gamma}$,
while for $C^{(\prime)}_{g}$, one only has the $N_c$ factor when a gluon
attaches to the internal gluino line,
which can be easily understood by using 't~Hooft's double line notation.
This is also true for the $F_1$ vertex term.
However, the chiral enhancement factor $m_{\tilde g}/m_b$ 
accompanying $\delta_{LR,RL}$ is a unique
feature of the $F_2$-term.
The mechanism is generic and has been discussed in Ref. \cite{Fujikawa},  
but SUSY with LR squark mixings gives a beautiful example \cite{b2sp}.

%%%%%%%%%%%%%%%%%%%%%%%%%%%%%%%%%%%%%%%%%%%%%%%%%%%%%%%%%%%%%%%%%%%%%%%%%%%%
%%%%%%%
%Fig 4% 
%%%%%%%
\begin{figure}[t!]%t!  
\centerline{{\epsfxsize1.7 in \epsffile{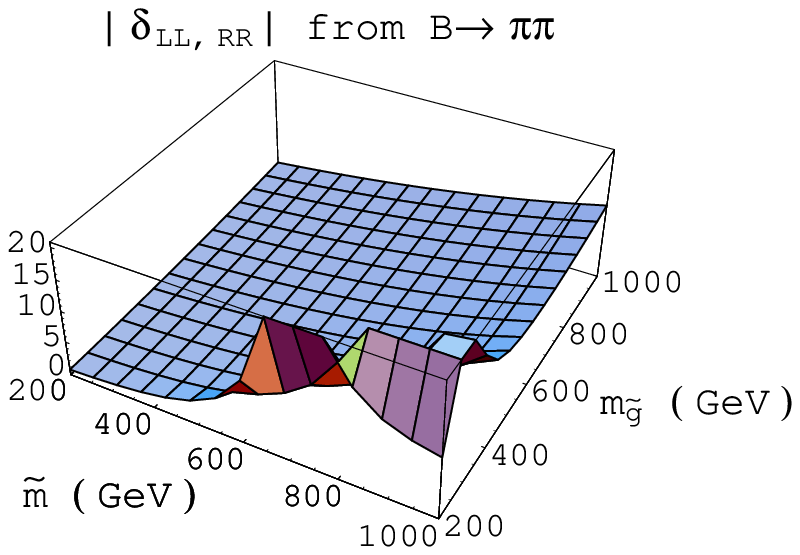}}  
            {\epsfxsize1.7 in \epsffile{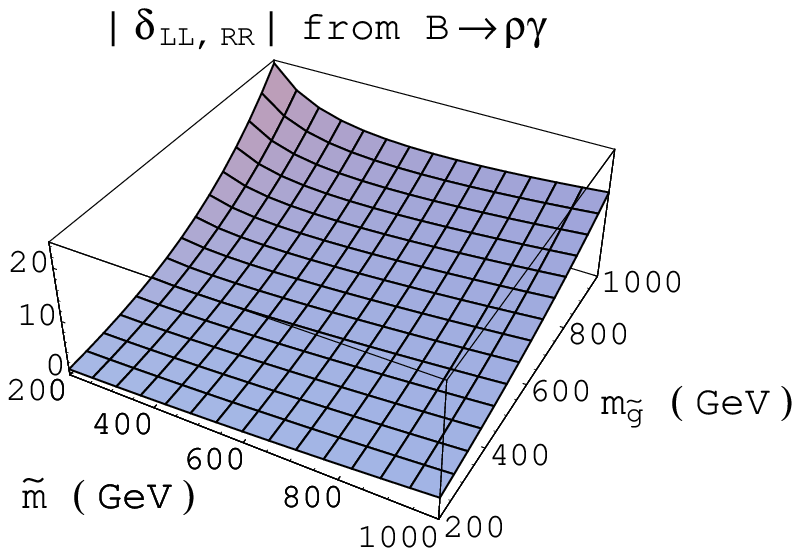}}}  
\smallskip  
\caption{Lower and upper limits on squark mixing angles $\delta_{LL,RR}$ obtained by 
(a) Br$^{\rm SUSY}(\pi^+\pi^-)/{\rm Br}^{\rm SM}(\pi^+\pi^-) > 10\%$,
(b) Br$^{\rm SUSY}(\rho^0\gamma)/{\rm Br}^{\rm SM}(\rho^0\gamma) < 4$,
respectively.
}
\label{fig:B2dllrr}  
\end{figure}  
%%%%%%%%%%%%%%%%%%%%%%%%%%%%%%%%%%%%%%%%%%%%%%%%%%%%%%%%%%%%%%%%%%%%%%%%%%%%

\begin{figure}[t!]
\centerline{{\epsfxsize1.7 in \epsffile{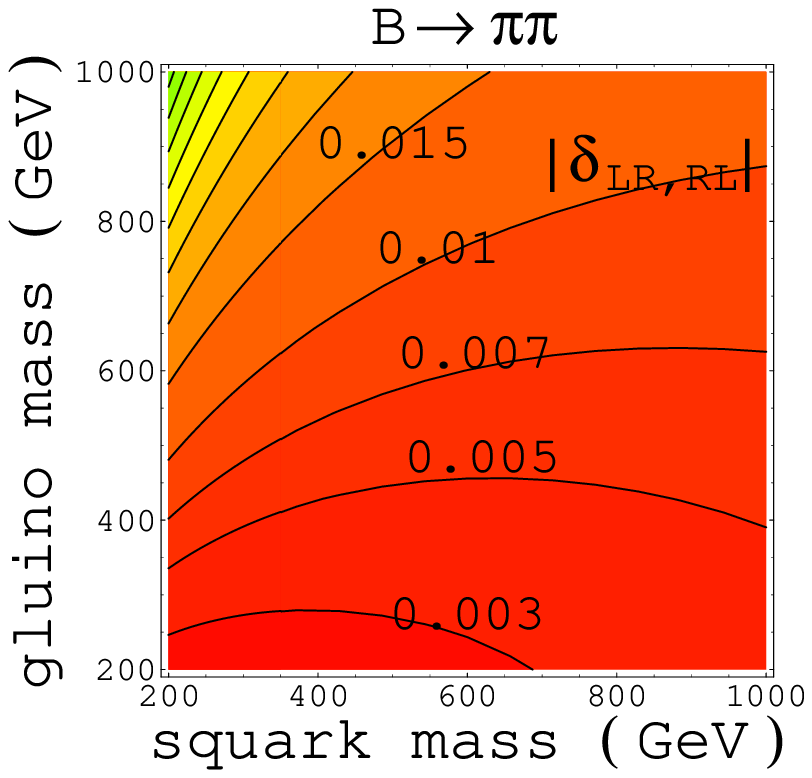}}  
            {\epsfxsize1.7 in \epsffile{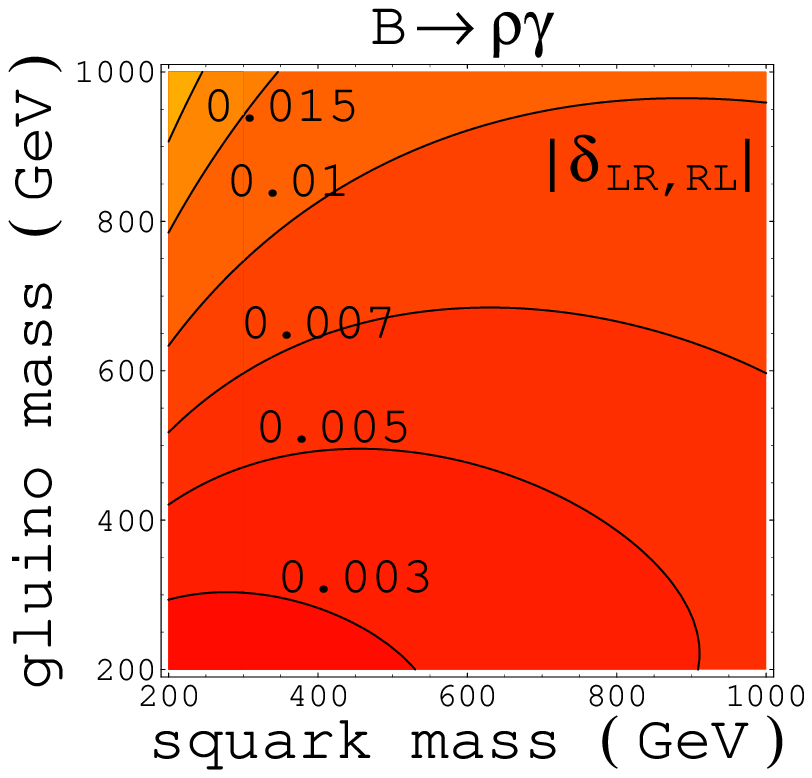}}}  
\smallskip  
\caption{Lower and upper limits on squark mixing angles $\delta_{LR,RL}$ obtained by 
(a) Br$^{\rm SUSY}(\pi^+\pi^-)/{\rm Br}^{\rm SM}(\pi^+\pi^-) > 10\%$,
(b) Br$^{\rm SUSY}(\rho^0\gamma)/{\rm Br}^{\rm SM}(\rho^0\gamma) < 4$,
respectively.}
\label{fig:B2dlrlr}  
\end{figure}

For a direct destructive interference to cut down by half the predicted SM rate,
one needs the SUSY amplitude to be 30\% of the SM amplitude.
This will be the minimum requirement on SUSY contribution.
In Figs. \ref{fig:B2dllrr}(a) and \ref{fig:B2dlrlr}(a) we show limits
on $|\delta_{LL,RR}|$ and $|\delta_{LR,RL}|$, respectively.
We require the SUSY contribution alone to give 
10\% of Br$^{\rm SM}(B\to\pi\pi)$,
corresponding to $\sim$30\% in amplitude.
If we change the required rate contribution by a factor $\kappa$, 
the values shown in the plots scale by a factor $\sqrt\kappa$. 
From Fig. \ref{fig:B2dllrr}(a), we see that the decay rate
is insensitive to left-left and right-right squark mixings,
which means gluino box and $\delta_{LL,RR}$ related gluino penguins
do not give large contributions.

%%%%%%%%%%%%%%%%%%%%%%%%%%%%%%%%%%%%%%%%%%%%%%%%%%%%%%%%%%%%%%%%%%%%%%%%%%%
%%%%%%%
%Fig 5% 
%%%%%%%
%%%%%%%%%%%%%%%%%%%%%%%%%%%%%%%%%%%%%%%%%%%%%%%%%%%%%%%%%%%%%%%%%%%%%%%%%%%

The SUSY contribution from gluino box and the $F_1$ term 
is dominated by $B_{1,2}(\xgq)$, $P(\xgq)$ and $g_1(\xgq)$.
For $\msq\ll\mgl$, $B_2(\xgq)$ is dominant 
and contributes through $c_6$, 
which is from gluino box containing $\tilde d_L$--$\tilde b_L$ and 
$\tilde q_R$ squark lines.
For $\mgl\ll\msq$, $P(\xgq)$ is dominant 
and contributes through $c_{4,6}$.
However, from Eqs. (\ref{amplitude}), (\ref{ci}), (\ref{asymptoticBP}), 
(\ref{c8New}) and (\ref{asymptoticg}),  we see that this $F_1$ term 
always receive cancellation from $B_1(\xgq)$ and $g_1(\xgq)$, 
which are not too small in this region. 
Therefore the rise in the lower right corner of Fig. \ref{fig:B2dllrr}(a) 
shows insensitivity to $\delta_{LL,RR}$ as a consequence of this
cancellation effect.

In Fig. \ref{fig:B2dlrlr}(a), 
we show the required $|\delta_{LR,RL}|$ to produce 
large enough SUSY contribution in $B\to\pi\pi$ decay.
Note that in Eq. (\ref{c8New}), we use a running $m_b(\mu_{\rm SUSY})$. 
%for example $m_b(m_t)=2.79$ GeV 
For most of the parameter space a less than 2\% mixing angle in 
left-right mixing is enough to generate such a large SUSY contribution.
The sensitivity is greatly enhanceded from the previous case 
due to the chiral enhancement factor $\mgl/m_b$.
Note that there is nothing peculiar about chiral enhancement.
It only reflects the chiral suppression of $C_{g,\gamma}$ in the SM
due to the V$-$A nature of weak interaction,
which need not be obeyed by interactions beyond the SM.

For left-right mixing case as shown in 
Fig. \ref{fig:B2dlrlr}(a),
we see that the SUSY contribution is larger for squark mass greater 
than gluino mass and vice versa.
For the case of heavy squark and light gluino ($\xgq\equiv\mgl^2/\msq^2<1$), 
the gluon preferably radiates off the gluino rather than the squark line,
as is clear from the behavior of $g_3(\xgq)$ in 
Eqs. (\ref{c8New}--\ref{asymptoticg}). 
Note that this is the one with $N_c$ enhancement 
and therefore gives larger contribution compared to $\msq<\mgl$ case, 
where the dominant diagrams do not have $N_c$ enhancement.
The SUSY contribution is dominated by 
$\delta_{LR,RL}\mgl g_3(\xgq)/(m_b\msq^2)$.
For $\msq\ll\mgl$, as shown in Eq. (\ref{asymptoticg}),
this term becomes $\delta_{LR,RL}\msq^2/(2\mgl^3 m_b)$ and is consistent
with the sharp rise in the upper left corner of Fig. \ref{fig:B2dlrlr}(a).
The SUSY contribution is small and 
insensitive to squark mixing angle in this region.
This is a generic feature for gluino penguin contributions, 
as one can see from Eq. (\ref{asymptoticg}) that all $g_i/\msq^2$ receive a 
$\msq^2/\mgl^4$ suppresion factor in this region. 
In the reversed case of $\mgl\ll\msq$,
there is only $1/\msq^2$ suppression, 
while $g_3(\xgq)$ contribution receive $\ln\xgq$ enhancement.

The gluino exchange induced photonic penguin is closely related to 
the gluonic penguin.
For example, they have similar chiral enhancement behavior
as well as asymptotic behavior.
Recently, the Belle Collaboration reports a 90\% upper limit on
Br$(B^0\to\rho^0\gamma)<1.06\times10^{-5}$\cite{Ushiroda}, 
nominally $\sim5$ times the SM prediction, but a factor of 2 above 
their previous result reported at ICHEP2000 \cite{Nakao}. 
We require the decay rate due to SUSY contribution alone to be smaller than 
4 times the SM prediction.
The bounds correspond to $|C^{(\prime)\rm SUSY}_\gamma|\sim 2\,|C^{\rm SM}_\gamma|$.
Note that in the LR case, one may have cancellation between SM and SUSY
$C_\gamma$.
For a direct cancellation, the bounds on $\delta_{LR}$ can be relaxed by 50\%. 
In Fig. \ref{fig:B2dllrr}(b) and \ref{fig:B2dlrlr}(b),
we show the limits on $|\delta_{LL,RR}|$ and $|\delta_{LR,RL}|$, respectively. 
Similar to the $B\to\pi\pi$ case, the decay rate is insensitive to
$|\delta_{LL,RR}|$, but very sensitive to $|\delta_{LR,RL}|$,
as expected.
We also see a sharp rise in the upper left corners of 
Fig. \ref{fig:B2dllrr}(b) and \ref{fig:B2dlrlr}(b), 
which are similar to \ref{fig:B2dlrlr}(a) 
and indicate that SUSY contributions are insensitive to the corresponding
squark mixings in that region as we have explained in the previous
paragraph.

We note that in most of the parameter space 
shown in Fig. \ref{fig:B2dlrlr}(b), 
$|\delta_{LR,RL}|$ are constrained to be less than 2\%. 
When compared to Fig. \ref{fig:B2dlrlr}(a), 
in most of the parameter space $|\delta_{LR,RL}|$ impacts
more on $B\to\rho\gamma$ than $B\to\pi\pi$. 
This can be easily understood by
noting that the former is a pure loop process while 
the latter is dominated by tree diagrams in SM. 
Therefore, it is easier for New Physics to affect the former process.

%%%%%%%%%%%%%%%%%%%%%%%%%%%%%%%%%%%%%%%%%%%%%%%%%%%%%%%%%%%%%%%%%%%%%%%%%%%
%%%%%%%
%Fig 6% 
%%%%%%%
\begin{figure}[t!]
\centerline{\,\,\epsfxsize3.28 in \epsffile{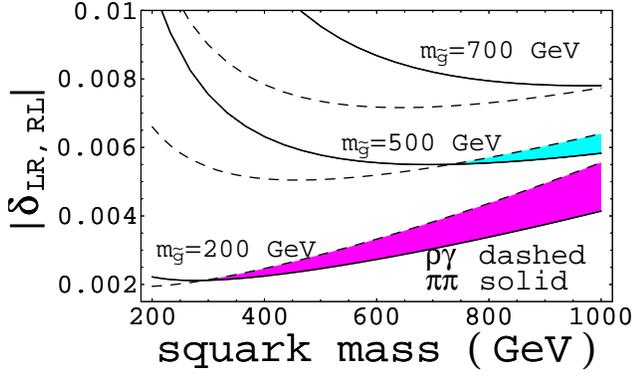}}  
\smallskip  
\caption{Dashed [solid] lines are upper [lower] bounds on squark mixing angles 
$\delta_{LR,RL}$ obtained by 
Br$^{\rm SUSY}(\rho^0\gamma)/$ Br$^{\rm SM}(\rho^0\gamma) < 4$
[Br$^{\rm SUSY}$ $(\pi^+\pi^-)/Br^{\rm SM}(\pi^+\pi^-)$ $> 10\%$]
with $m_{\tilde g}=200$, 500, 700 GeV, respectively.
Shaded regions are allowed parameter space.
}
\label{fig:rhogammapipi}  
\end{figure}  
%%%%%%%%%%%%%%%%%%%%%%%%%%%%%%%%%%%%%%%%%%%%%%%%%%%%%%%%%%%%%%%%%%%%%%%%%%%

It is quite interesting that, for the parameter space of
$\tilde m\sim 300$--$1000$ GeV and $m_{\tilde g}\leq700$ GeV, 
with mixing angle $\sim 0.2\%$--0.8\%, the model gives 
sizable contribution to $B\to\pi\pi$ decay that can account for the
smallness of the rate, but still satisfy the $B\to\rho\gamma$ constraint.
As shown in Fig. \ref{fig:rhogammapipi}, 
where the dashed (solid) lines correspond to upper (lower) limits on
$|\delta_{LR,RL}|$ from $B\to\rho\gamma\,(\pi\pi)$, 
with $m_{\tilde g}=200,$ 500, 700 GeV, respectively.
The shaded regions are the allowed parameter space for given $\mgl$.
For $m_{\tilde g}>700$ GeV we need $\widetilde m>1$ TeV to have allowed region,
which is beyond the plot. 
In addition, one can also use a smaller $\langle q^2 \rangle$, such as
$m^2_b/4$, to enhance the color dipole contribution and 
thus reduce the limit by 33\% and enlarge the overlapping parameter space
between Figs. \ref{fig:B2dlrlr}(a) and \ref{fig:B2dlrlr}(b).  
The existence of this overlap region is closely related to the behavior of 
$g_i(\xgq)$. 
From Eqs. (\ref{c7New}) and (\ref{c8New}), it is clear that for left-right mixing, 
$C^{\prime}_\gamma$$\,\propto\,$$\mgl/m_b\,g_4(\xgq)$, 
while $C^{\prime}_g$$\,\propto\,$$\mgl/m_b\,g_3(\xgq)$.
For $\xgq\ll1$, $g_3(\xgq)/g_4(\xgq)\to6|\ln\xgq|$ can be rather sizable.
Therefore, gluino penguin can give larger contribution 
in $b\to d g$ than in $b\to d\gamma$ process.
 
%%%%%%%%%%%%%%%%%%%%%%%%%%%%%%%%%%%%%%%%%%%%%%%%%%%%%%%%%%%%%%%%%%%%%%%%%%%
%%%%%%%
%Fig 7% 
%%%%%%%
\begin{figure}[b!]
\centerline{{\epsfxsize1.8 in \epsffile{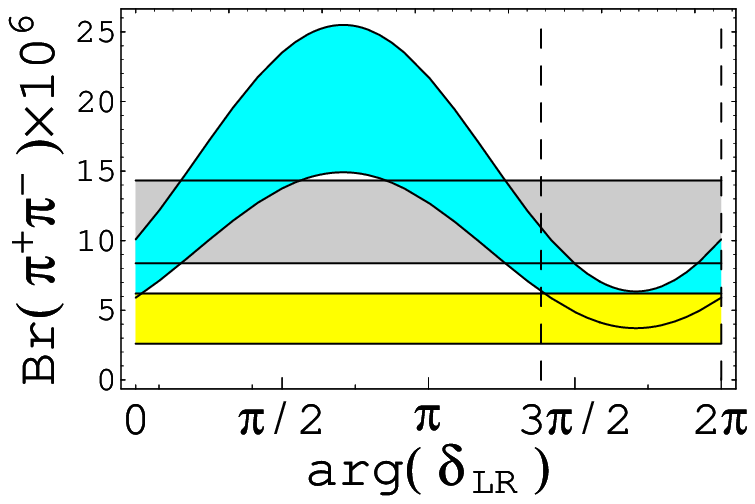}}  
            {\hspace{-0.2cm}\epsfxsize1.8 in \epsffile{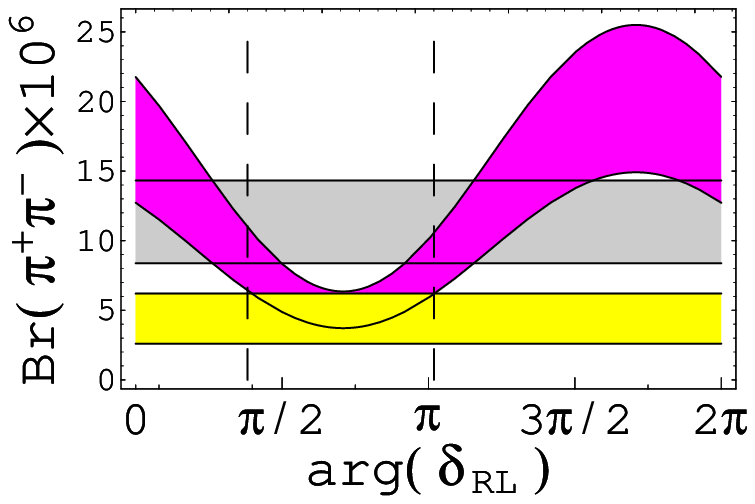}}}  
\smallskip  
\caption{ Br$(B\to\pi\pi)$ obtained by using $\msq=800$ GeV, $\mgl=200$ GeV,
and
(a) $|\delta_{LR}|=0.0035$, 
(b) $|\delta_{RL}|=0.0035$,
respectively.
The upper band corresponds to the SM prediction, while the lower band corresponds to
the experimental result with $2\sigma$ error range.}
\label{fig:B2pipiexam}  
\end{figure}  
%%%%%%%%%%%%%%%%%%%%%%%%%%%%%%%%%%%%%%%%%%%%%%%%%%%%%%%%%%%%%%%%%%%%%%%%%%%

%%%%%%%%%%%%%%%%%%%%%%%%%%%%%%%%%%%%%%%%%%%%%%%%%%%%%%%%%%%%%%%%%%%%%%%%%%%
%%%%%%%
%Fig 8% 
%%%%%%%
\begin{figure}[t!]
\centerline{{\epsfxsize1.86 in \epsffile{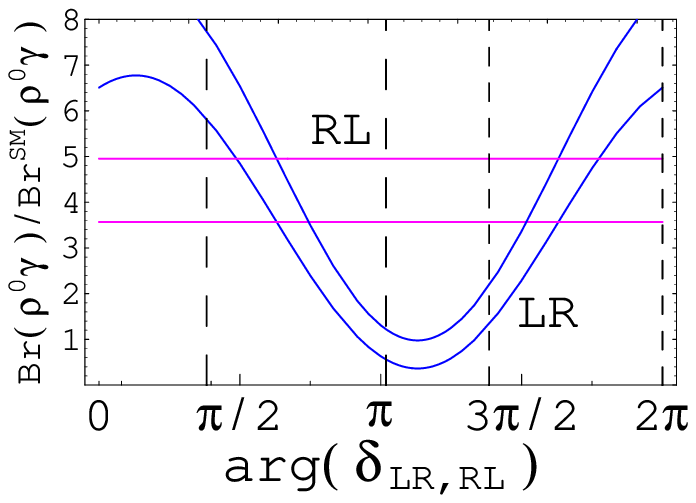}}  
            {\hspace{-0.5cm}\epsfxsize1.8  in \epsffile{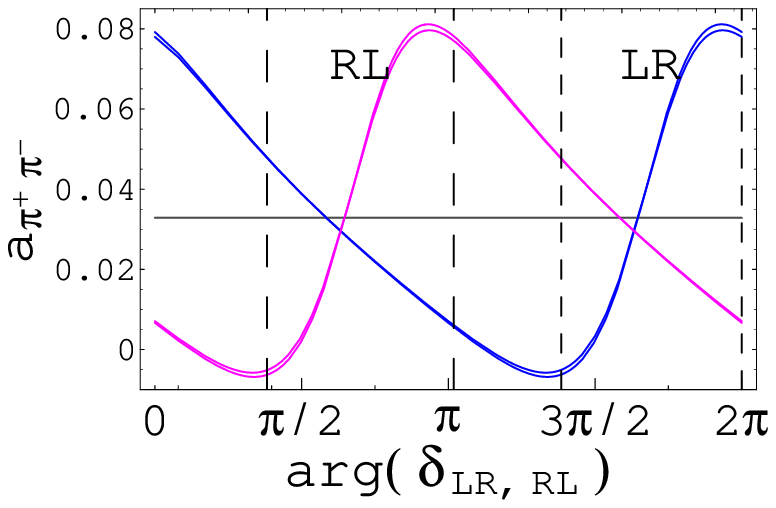}}}  
\smallskip  
\caption{ 
(a) Upper (lower) lines corespond to
    Br$(\rho^0\gamma)/$ Br$^{\rm SM}(\rho^0\gamma)$ with $\mgl=$200 (500) GeV, 
    $\msq$=800 (900) GeV, $|\delta_{LR,RL}|$=0.0035 (0.006). 
(b) The asymmetry in $B\to\pi^+\pi^-$ with same parameter space as case (a).
}
\label{fig:B2rhogammaacp}
\end{figure}  
%%%%%%%%%%%%%%%%%%%%%%%%%%%%%%%%%%%%%%%%%%%%%%%%%%%%%%%%%%%%%%%%%%%%%%%%%%%

For illustration, we pick some points from Fig. \ref{fig:rhogammapipi} and study
the impact of SUSY contributions on Br$(\pi^+\pi^-)$ and Br$(\rho^0\gamma)$. 
In Fig. \ref{fig:B2pipiexam},
we show Br$(\pi^+\pi^-)$ obtained by using $\mgl=200$ GeV, $\msq=800$ GeV,
and
(a) $|\delta_{LR}|=0.0035$, $\delta_{RL}=\delta_{LL,RR}=0$,
(b) $|\delta_{RL}|=0.0035$, $\delta_{LR}=\delta_{LL,RR}=0$,
respectively.
The upper band corresponds to the SM prediction, while the lower band corresponds to
the averaged experimental result Br$(\pi^+\pi^-)=4.4\pm0.9$ with $2\sigma$ 
error range. 
With arg($\delta_{LR,RL}$) within the dashed lines,
i.e. arg($\delta_{LR})\sim 4.3$--2$\pi$, arg($\delta_{RL})\sim 1.2$--3.2,
Br$(\pi^+\pi^-)$ can be brought down by SUSY contributions to experimental range.
The strength factor of $a_{\rho\gamma}$ \cite{b2sp,Atwood} 
\begin{equation}
\sin 2\theta\equiv {2|C_\gamma C^\prime_\gamma|
                    \over
                     |C_\gamma|^2+|C^\prime_\gamma|^2},
\end{equation}
can be as large as 90\% in this case.
The measurability of the asymmetry in $B\to\rho\gamma$ decay is better
than in $B\to K^*\gamma$ since it readily provides vertex
information\cite{Atwood}.

It is clear from Fig. \ref{fig:rhogammapipi} that we might as well take 
$\mgl=500$ GeV, $\msq=900$ and $\delta_{LR,RL}=0.006$.
The previous case corresponds to $\xgq=0.06$, while in this case we have
larger $\xgq=0.31$.
These two cases also represent other cases with
similar $\xgq$, while $\msq$ need not be that heavy.
The figure for Br$(\pi^+\pi^-)$ are almost identical to Fig. \ref{fig:B2pipiexam}.
However, as we show in Fig. \ref{fig:B2rhogammaacp}(a), the latter case has greater
Br$(\rho^0\gamma)$.
Note that RL case is insensitive to arg$(\delta_{RL})$, since the rate is 
proportional to $|C_\gamma|^2+|C^\prime_\gamma|^2$.
In this case, Br$(\rho^0\gamma)$=
Br$^{\rm SM}(\rho^0\gamma)$+Br$^{\rm SUSY}(\rho^0\gamma)$ is within 5 times of 
SM rate as required by Fig. \ref{fig:B2dlrlr} and Fig. \ref{fig:rhogammapipi}. 
All $B\to\pi\pi$ favored range, arg$(\delta_{RL})\sim 1.2$ -- 3.2 are also allowed
by the $B\to\rho\gamma$ constraint.
The $\sin 2\theta$ is 80\%.
For the LR case, the induced $C_\gamma^{\rm SUSY}$ may have constructive or
destructive interference with $C_\gamma^{\rm SM}$ as arg($\delta_{LR}$) changes.
Within the arg($\delta_{LR})\sim 4.3$--$2\pi$ range, allowed by $B\to\pi\pi$ rate,
there are quite some parameter space
to satisfy the $B\to\rho\gamma$ constraint.
The rate can be close to the SM expectation.
In Fig. \ref{fig:B2rhogammaacp}(b), we show the asymmetry, $a_{\pi^+\pi^-}$.
Note that the SUSY prediction for $a_{\pi^+\pi^-}$ from the previous two paramter 
points are close as in the Br($\pi^+\pi^-$) case.
In generalized factorization, $a^{\rm SM}_{\pi^+\pi^-}\sim 3.3\%$ with our 
input parameter,
a smaller $\langle q^2\rangle$ will have a slightly larger asymmetry.
With SUSY, $a_{\pi^+\pi^-}$ can be ranging within $-1$\% to 8\%.
Note that in QCD factorization approach including weak annihilation contribution
$a^{\rm SM}_{\pi^+\pi^-}\sim -5\%$ to 15\%, 
for $\phi_3\sim 60^\circ$ \cite{Beneke:2001ev}.
It is difficult to distingiush SUSY contributions from the SM prediction from
$a_{\pi^+\pi^-}$.

\section{Discussion}  

Left-left and/or right-right \cite{ChuaHou} 
$\tilde d$--$\tilde b$ mixings with few \% to few 10\% mixing angle
can generate large enough contribution to $B_d$-mixing that 
could reduce $a_{J/\psi K_S}$ from its SM value.
As shown in Fig. \ref{fig:B2dllrr}(b), such squark mixing angles 
are safe from $B\to\rho\gamma$ constraint.
However, as shown in Fig. \ref{fig:B2dllrr}(a), 
one needs large mixing with sizable mass splitting
to affect $B\to\pi\pi$ decay rate in this case.
Such a large mixing is already ruled out by the experimental measurement
of $\Delta m_{B_d}$, as shown in Figs. \ref{fig:Bdllrr}(a) and \ref{fig:Bdllrr}(b),
unless one fine tunes the parameter space to be very close to 
$\xgq=2.43$, and turn off left-left or right-right mixings.
In other words, one needs high degree of fine tuning to account for both
low $\sin2\phi_1$ and $B\to\pi\pi$ decay rate with LL or RR mixings.
It is much easier to compete with the SM box diagram 
and modify $\sin2\phi_1$ 
than to compete with tree dominated $B\to\pi\pi$ decay.

Alternatively,
left-right and/or right-left $\tilde d$--$\tilde b$ mixings 
with few \% mixing angles could also give 
sizable contribution to $B_d$ mixing. 
However, because of the amplification effect of chiral enhancement, 
the size of this mixing angle is severely constrained by $B\to\rho\gamma$,
to be less than 2\% in most of the parameter space
given in  Fig. \ref{fig:B2dlrlr}(b). 
It cannot be the source that gives
sizable contribution to $B_d$ mixing, 
as one can tell by comparing Figs. \ref{fig:Bdlrlr} and
\ref{fig:B2dlrlr}(b). 
It is interesting that, as noted already in the previous section, 
there is parameter space where $\mgl$ is suitably light and the
mixing angle $\delta_{LR,RL}$ is less than 1\%,  
where the model gives sizable contribution to $B\to\pi\pi$ decay 
without violating $B\to\rho\gamma$ constraint. 
In other words, we need left-right and/or right-left mixings 
rather than left-left and/or right-right mixings 
to affect $B\to\pi\pi$ decay. 
Thus, if the smallness of Br$(B\to\pi\pi)$ is due to SUSY,
it is likely that one will have large effects in $b\to d\gamma$,
including rate enhancement and mixing induced asymmetry\cite{Atwood}, 
which can be easily close to 100\%. 

In Section III, we used $\phi_3=65^\circ$, which is a CKM fit like value,
since SUSY contribution to $B\to\pi\pi$ decay is uncorrelated with
the SUSY contribution to $B_d$ mixing. 
The CKM fit may still be
viable and need not support large $\phi_3$ as a solution of low
$B\to\pi\pi$ rate. 

It is clear that we still need correct interference patterns, 
i.e. correct SUSY phases, to reduce $a_{J/\psi K_S}$ and
Br$(B^0\to\pi^+\pi^-)$. 
Since the effects arise from different squark
mixing sources, one can always find separate SUSY phases to achieve this.
As we show in Section II and Section III,
we may have accessible allowed region on SUSY phases.

\section{Conclusions}

We have shown that it is possible for SUSY models 
to account for the smaller $\sin2\phi_1$ and Br$(B\to\pi\pi)$ values
that seem to be emerging from the $B$ Factories.
However, they would have to come from different flavor mixing sources.
The smallness of $\sin2\phi_1$ is most likely arising from left-left 
(right-right) squark mixing,
while the deficit in Br$(B\to\pi\pi)$ is 
most likely due to left-right squark mixings.
The two are basically uncorrelated.

Because of similarity in chiral enhancement, 
the loop induced $b\to d\gamma$ process is even more sensitive to
$\delta_{LR,RL}$ than the tree dominated $B\to\pi\pi$ decay.
Therefore, if SUSY affects the latter,
the effects in the former would be even more prominent.  
We emphasize that 
$B\to\rho\gamma$ could be considerably larger than expected in SM
if the smallness of $B\to\pi\pi$ rate is in part due to SUSY.

\vskip 0.3cm  
\noindent{\bf Acknowledgement}.\ \  
A.A. is on leave of absence from Department of Mathematics FSTT, 
P.O. Box 416, Tangier, Morocco. 
This work is supported in part by
NSC-89-2112-M-002-063, 
NSC-89-2811-M-002-0086 and 0129, 
the MOE CosPA Project, 
and the BCP Topical Program of NCTS.

%\pagenumbering{}                                                       
%%%%%%%                                                                 
%Fig A%
%%%%%%%
%\begin{figure}[htb]
%\centerline{{\epsfxsize12 cm \epsffile{file.eps}}}  
%\smallskip  
%\caption {}  
%\label{fig:}  
%\end{figure} 
%%%%%%%
%Fig B%
%%%%%%%
%\begin{figure}[htb]  
%  
%\centerline{\DESepsf(file.eps width 12cm)}  
%\smallskip  
%  
%\caption{}  
%\label{fig:}
%  
%\end{figure}

\end{document}